\documentstyle[prl,aps]{revtex}
\begin{document}
\twocolumn[
\hsize\textwidth\columnwidth\hsize\csname@twocolumnfalse\endcsname
\draft
\title{ Haldane's Fractional Statistics  and
the Lowest Landau Level on a Torus}
\author{Ansar Fayyazuddin$^{1}$ and Dingping Li$^{2,3}$}
\address{$^1$NORDITA, Blegdamsvej 17, DK-2100 Copenhagen {\O}, Denmark}
\address{$^2$Dipartimento di Fisica and Sezione I.N.F.N.,
Universit\'a di Perugia, Via A. Pascoli, 
I-06100 Perugia, Italy}
\address{$^3$Division de Physique Th\'eorique,
IPN, Orsay Fr-91406, France}
\date{\today}
\maketitle
\begin{abstract}
The lowest Landau level on a torus is studied. The dimension of
the many-body Hilbert space is 
obtained and is found to be different from
the formula  given by Haldane. 
Our result can be tested in numerical investigations of 
the low-energy spectrum of
fractional quantum Hall states on a torus.
\end{abstract}

\pacs{PACS numbers: 73.20.Dx, 03.65.Fd, 05.30.-d, 73.20.mf.}  
] 

Identical particles with statistics interpolating between
Bose and Fermi can exist in two dimensions
\cite{lei}. Particles which obey fractional statistics
(neither bosonic nor fermionic) are called anyons.
Anyon physics has been applied to the theory of the
fractional quantum Hall effect (FQHE) \cite{laugh,pg}.
The condensation of the anyonic quasiparticles \cite{halperin,arovas}
in Laughlin states gives rise to hierarchical states in the FQHE 
\cite{halperin,haldane1}.

Recently,  Haldane proposed \cite{haldane}
a new definition of fractional statistics
by generalizing  the Pauli exclusion principle  
(we will call it Haldane's statistics or HS).
The definition of HS is based on a Hilbert space counting argument
(see Refs.\ \cite{ouvry,wu,isakov} for  further developments).
In the framework of HS  the monodromy properties of 
wavefunctions (which are usually used in the definition of statistics) 
are not used  to define the statistics and  the definition of 
statistics is independent of the dimension of space.

Following \cite{haldane}, we consider a $N$ body system in a finite space, 
and fix $N-1$ particles among them.
Analyze the Hamiltonian of the remaining particle 
assuming  that the dimension 
$d$ of the Hilbert space for this particle
is finite and independent
of which particle has been chosen.
The  parameter $g$ of HS is by definition 
\begin{equation}
\Delta d =-g \Delta N.
\label{core}
\end{equation}
For bosons, $g=0$, and for fermions, $g=1$.
For $g$ neither $0$ nor $1$, we say that particles obey
fractional statistics. If Eq.\ (\ref{core}) is satisfied, we say that 
particles satisfy  HS.  The size of the Hilbert 
space  of the many-body system
at fixed $N$ has been argued to be equal to \cite{haldane}
\begin{equation}
{(d+N-1)! \over (N)!(d-1)!} \, \, .
\label{size}
\end{equation}
One of the advantages of the concept of HS is that
for bosons or fermions the number of N-body states
among  accessible states  is given by the unified Eq.\ (\ref{size}). 
In various examples of HS, for example, anyons
in a very strong magnetic field on   a sphere,
Haldane-Shastry Spin Chain model etc.,
Eq.\ (\ref{size})  was found to be satisfied. 
One would guess it may be true in any example of HS.

In Refs.\ \cite{he,joh}, the emphasis was put on the Hilbert space
of the low-energy sector of fractional quantum Hall (FQH) states
{\bf {\it on a sphere}},  
in the presence of quasielectron (QE) or quasihole (QH) excitations.
On the other hand, the energy spectrum of a few electrons can
be calculated numerically. A low-energy sector was found which
is well separated from the groundstate and corresponds 
to Hall states with QE or QH excitations. The dimension
of the Hilbert space of such states  
predicted in Ref.\ \cite{haldane} by using Eq.\ (\ref{size}),   
for example, in the presence of QH excitations, is
${(N_e+N_q)!/ [(N_e)!(N_q)!]}$,
which is in agreement with 
the number of states in the low-energy sector (NSLES)
found in numerical studies.  Furthermore, anyons on the sphere  with 
hard core boundary conditions 
interacting with a strong magnetic field 
were considered in \cite{sphere}, and 
the dimension of the Hilbert space of
many-body states (DHMS)   was found  
to be given  by Eq.\ (\ref{size}), in agreement with   \cite{he,joh}.

In this letter, we consider anyons in 
a strong magnetic field on a torus,
thus only the lowest Landau level (LLL) states. 
We explicitly work out the formula of the DHMS,
which is the  number of possible many-body states in the LLL.
We note that the concept of HS can also be applied to our case
and  expect  the formula of the DHMS  to be 
given by Eq.\ (\ref{size}) as in the case of spherical
geometry. However,   the DHMS found in this case  is 
{\bf {\it not}} given by Eq.\ (\ref{size}).
By using the results of this letter 
we also obtain the  NSLES
of  FQH states on a torus
by using Halperin's theory of hierarchical states.  
Our formula can be tested in numerical calculations
on a torus, similar to
Ref.\ \cite{joh} where the calculation was performed on a sphere.
We conclude that when the topology is changed to that of a torus,  
the  prediction made by HS does not 
tally with what we find 
for anyons in the LLL on a torus.
One may guess that Eq.\ (\ref{size}) is
valid only when there are no topological obstructions for
constructing single component wavefunctions. 
The nontrivial topology of tori
forces the wavefunctions of anyons to be multi-component,
in contrast to the  case  of  spheres
\cite{ein,braid,iengo,ansar,hosot}.

The NSLES  of FQH states  on a {\bf {\it sphere}} was obtained in
Jain's composite fermion theory of the FQHE  \cite{jain}, 
and the result was found to be identical to 
the one predicted in Halperin's hierarchical  
theory using the HS concept.
However we do not know how to use Jain's theory to
obtain the NSLES of FQH states on a torus because 
of difficulties in the construction of 
Jain's wavefunctions  on a torus with the correct degeneracy.
FQH states have  degenerate ground states 
on a torus \cite{degen}, which can be seen in 
numerical calculations \cite{suwp}. 
We note that one can construct Halperin's hierarchical 
wavefunctions on a torus with the correct degeneracy \cite{torusli}. 

We begin by solving the DHMS of anyons 
in the LLL on a torus  at a specific filling.
Following the notation in Ref.\ \cite{torusli},
the Hamiltonian of a quasihole in singular gauge 
 \cite{iengo,ansar,hosot} is
$H={1\over 2m} [{(p_x-e_qBy)}^2+{(p_y)}^2]$,
where $B$ is the magnetic field and
$e_q$ is the charge of the particle. 
Let $\Phi$ be the magnetic flux seen by the particle,
$\Phi =\tau_2e_qB$.
We take the statistical parameter  $\pi \alpha =\pi k/s$,
where $k,s$ are positive coprime integers. 
The statistical phase is then $e^{i\alpha \pi}$.
$\alpha$ must be a rational number on a torus and
$\Phi - \alpha N=\Phi^{\prime}$ must be an integer according to
the Dirac quantization condition. 
Thus we are considering the case 
where the statistical flux is opposed
to the flux of the external magnetic field.  
The situation where the two fluxes are
parallel with hard core boundary conditions 
imposed as in Refs.\ \cite{he,joh,sphere} will
be considered in detail elsewhere and
a short discussion of such a situation in the case of QE excitations
can be found at the end of this letter.
Although it is usual to restrict $\mid\alpha\mid <1$ because
the statistical phase $e^{i\alpha \pi}$ 
depends only on $\alpha\bmod 2$, 
we  allow $\alpha >0$ to be unrestricted
as in Ref.\ \cite{joh}.  Here $\alpha $ is related to 
the amount of statistical flux bound to each particle.   
The wavefunctions are represented by $s$-component vectors.
In our case, the wavefunctions  can be written as
\begin{eqnarray}
{ \Psi}(z_{i} )_{n}
&=& \exp (-{\pi \Phi 
\sum_i y^2_{i} \over 
\tau_2})  \prod_{i<j} 
{[\theta_3(z_{i}- z_j |\tau)]}^{k/s}
F(z_{i})_{n}   
\label{3bbell} 
\end{eqnarray}
where $n=1, \cdots , s$ is the component index
of the wavefunctions (all numbers appearing above and below
are positive if not otherwise specified).

We  can separate the center of mass 
coordinate part of the wavefunctions
from the relative part in the case of 
$\Phi^{\prime}=pN$ where $p$ is an  integer
by generalizing the arguments of Ref.\ \cite{degent}.
As $\Phi /N=(k/s)+p$,
the functions  $F(z_{i})_{n}$ are given as
\begin{eqnarray}
F(z_{i})_{n} =  \theta {a_{m, n}  \brack b} 
(\sum_{i} z_{i} e_2
|e_1,\tau) F^{\prime}
 \, .   
\label{3bbe} 
\end{eqnarray}
$F^{\prime}$ depends on the relative holomorphic coordinates
($z_i-z_j$) and satisfies certain translational 
properties around the handles (or the nontrivial homology cycles)
on the torus.
In Eq.\ (\ref{3bbe}),  the notation is
\begin{eqnarray}
& &(e_1)^2=s(k+ps)\, ,(e_2)^2=p+(k/s) \, , \nonumber \\
& &a_{m, n}=
a^{\ast}_{m,n}e^{\ast}_1 \, ,
b=b^{\ast}e^{\ast}_1 \, ;  e^{\ast}_1={1 \over e_1}, 
e^{\ast}_2={1\over e_2} \, ,
\nonumber \\ 
& &a^{\ast}_{m,n}=a_0
+ m s +n(k+ps) \, ,  
b^{\ast}_i=b_0 \, ,\nonumber \\
& &m=1, \cdots ,k+ps \, ,
n=1, \cdots , s 
\label{3bbf} 
\end{eqnarray}
$a_0,b_0$ are fixed by boundary conditions and
$m$ is the index of the center-coordinate
degeneracy (CCD). 
When $p$ is an even integer,  
for the wavefunctions of  Laughlin type,
$F^{\prime}$  is equal to  
$F^{\prime}(p)_{lau}= \prod_{i<j} 
{[\theta_3(z_{i}- z_j |\tau)]}^p$. 
therefore in this case,
the degeneracy of the wavefunctions
is equal to the CCD: $k+ps$ \cite{torusli}. 

To obtain the DHMS of the model
when $\Phi^{\prime}=pN$ and $p$ is an even integer,
we find the number of all possible wavefunctions, 
or possible $F^{\prime}$ (or possible $F_n$),  
not necessarily restricted to be of Laughlin type. 
The wavefunctions of the center coordinate part shall 
remain unchanged with respect to different solutions of
$F^{\prime}$ \cite{degent}.

To work out $D_r$, the number of all possible  $F^{\prime}$, 
it is not necessary to get  explicit solutions 
for all possible $F^{\prime}$.
Considering $N$ bosonic particles  and taking 
the magnetic flux quanta $\Phi$ equal to $\Phi_p=pN$,
the wavefunctions of this model are   given by
Eq.\ (\ref{3bbe}), with $\Phi =pN\,$,
$(e_1)^2=p \, , (e_2)^2=1\,$, $\alpha =0$ and 
$m=1, \cdots , p \, , n=1$.
One can  show that $F^{\prime}_{bo}$,
the wavefunction of the relative part in the bosonic model,
satisfies the same translational properties 
and has the same dependence 
on the relative holomorphic coordinates
as $F^{\prime}$ in the anyon model.
Thus we have the same solutions for
$F^{\prime}$ and $F^{\prime}_{bo}$ 
according to Algebraic Geometry, 
this implies that the number of possible $F^{\prime}$, $D_r$,
is equal to the number of  possible $F^{\prime}_{bo}$.  
The DHMS 
(we recall that the DHMS is the number of all possible states)
of the bosonic model is 
$D_{bo}={(\Phi_p +N-1)! / [(\Phi_p -1)! (N)!]}$
because the number of possible states for
a single particle is equal to $\Phi_p$.
Therefore $D_r$  is equal to 
$D_{bo}$ divided by the  CCD of the wavefunctions,  
which is equal to $p$ here.
This leads to $D_{r}={(1/ p)} D_{bo}
={(\Phi_p +N-1)! / [(\Phi_p )! (N-1)!]}$. 
Thus the DHMS of the anyon model is equal to
$D=D_r \times (k+ps)$
where $(k+ps)$ is the CCD  of the  model.
By using the relation $\Phi^{\prime} =\Phi_p=pN$,
we have  $N(k+ps)=Nk+\Phi^{\prime} s$.
Inserting this relation in the formula for $D$ we obtained above,  
$D$ can be rewritten as
\begin{eqnarray}
D=(Nk+\Phi^{\prime} s)
{(\Phi^{\prime} +N-1)! \over (\Phi^{\prime} )! (N)!}  \, \, \, .
\label{dhms}
\end{eqnarray}
In the discussion above
we obtained the DHMS of anyons in the LLL 
on a torus when  $\Phi^{\prime}=\Phi_p =pN$ and $p$ is
an even integer.
Eq.\ (\ref{dhms}) depends only on $\Phi^{\prime}\,$,
$\alpha$ ($\alpha =k/s$) or  $\Phi$ and $\alpha$.
There is no explicit
dependence on the parameter $p$ in the final formula
although we used the relation $\Phi^{\prime} =pN$ 
in the derivation of the formula.
 
We now generalize the discussion 
to the case of arbitrary  integer  $\Phi^{\prime}$
and show that formula\ (\ref{dhms}) still holds.
For any integer $\Phi^{\prime}$, one can write
$\Phi^{\prime}=[k^{\prime}/s^{\prime}]N$ where 
$k^{\prime},s^{\prime}$ are coprime integers.

We first consider N bosonic particles with flux
$\Phi_{bo}=\Phi^{\prime}=[k^{\prime}/s^{\prime}]N$. 
The wavefunctions are given by Eq.\ (\ref{3bbell})
with $\Phi=\Phi_{bo}$.
The wavefunctions of the center coordinate part in
$F(z_i)_{bo}$  
are given by 
 $\theta {a_{m, l}  \brack b} 
(\sum_{i} z_{i} e_2 |e_1,\tau)$ 
(these can be obtained, for example,  by following 
Ref.\ \cite{degent}),   where
$(e_1)^2=k^{\prime}s^{\prime}
\, ,(e_2)^2= (k^{\prime}/s^{\prime}) \,$,
$a^{\ast}_{m,l}=a_0+
 m s^{\prime} +lk^{\prime}\, $  and
$m=1, \cdots , k^{\prime}\, ,
l=1, \cdots , s^{\prime} $. 
The   CCD of the model is equal to $k^{\prime}$.
Because $l$ is not restricted to be one, 
when one coordinate is moved around the handles,
the index $l$  will be changed.
One may say that the wavefunctions are multi-component.
However, this cannot be correct 
as the wavefunctions of bosonic and fermionic particles
are single-component.
This apparent paradox arises due to the fact that
factorization of the wavefunction into relative and center parts
is possible {\it only} when $\Phi / N$ is an integer due
to obstructions caused by 
translation properties of the wavefunctions.
In the present situation,  
$F(z_i)_{bo}$ should be expanded as
$\sum_l \theta {a_{m, l}  \brack b} 
(\sum_{i} z_{i} e_2 |e_1,\tau)F^{\prime \, bo}_l$.
The wavefunctions of the relative part 
(denoted as $F^{\prime \, bo}_l$ here )
are also multi-components and satisfy certain translational
properties.  
When one coordinate is moved around the handles,
the index $l$  inside the summation
will be changed to another number.
But the final wavefunctions are single-component due 
to the contraction  of the index $l$.

Now we consider the anyon model with 
$\Phi^{\prime}=(k^{\prime}/s^{\prime})N$. 
As $\Phi /N=(k/s)+(k^{\prime}/s^{\prime})=
(ks^{\prime}+sk^{\prime})/(ss^{\prime})$,
the wavefunctions of the center coordinate part in $F(z_i)$  
are given by the functions
 $\theta {a_{m,n,l}  \brack b} 
(\sum_{i} z_{i} e_2 |e_1,\tau)$  with
$(e_1)^2=s^{\prime}s (ks^{\prime}+sk^{\prime})
\, ,(e_2)^2= (k/s)+
(k^{\prime}/s^{\prime}) \,$,
 $a^{\ast}_{m,n,l}=a_0
+ m ss^{\prime} +(ls+n)
(ks^{\prime}+sk^{\prime}) \,$  and
$m=1, \cdots , (ks^{\prime}+sk^{\prime})\, ,
n=1, \cdots ,s \, , l=1, \cdots , s^{\prime} $.
The CCD is equal to $ ks^{\prime}+sk^{\prime}$.
For wavefunctions of the relative part in $F(z_i)$
(denoted as $F^{\prime}$ here),
one can show that $F^{\prime}$ of the bosonic model
and $F^{\prime}_{bo}$ of 
the anyon model satisfy the same translational
properties and have the same dependence on  holomorphic coordinates.
Thus the possible $F^{\prime}$
are given by  $s^{\prime}$-dimensional vectors
$F^{\prime}_l$ with $l=1, \cdots , s^{\prime} $,
similar to the case  of the bosonic model. 
By the same reasoning as for the bosonic model, 
we can expand $F$  as 
$\sum_l\theta {a_{m,n,l}  \brack b} 
(\sum_{i} z_{i} e_2 |e_1,\tau)F^{\prime}_l$.
The final wavefunctions have
$s$-components and the component index is $n$.

Since the vectors $F^{\prime}_l$ satisfy
the same translational
properties as the vectors $F^{\prime \, bo}_l$ and
depend only on holomorphic coordinates,  
the number of possible $F^{\prime}_l$ is equal to
the number of possible $F^{\prime \, bo}_l$.
Let $D$ be the DHMS of the anyon model
and $D_{bo}$ be that of the bosonic model. 
As the  wavefunctions of the relative part for both models 
have the same degeneracy,  
$D$ divided by $D_{bo}$ 
is equal to the CCD of the anyon model
divided by that of the bosonic model.
So $D/D_{bo}= ( ks^{\prime}+sk^{\prime})/k^{\prime}$.
Since $D_{bo}={(\Phi^{\prime} +N-1)! / [(\Phi^{\prime} -1)! (N)!]}$,
$D$ is given 
by Eq.\ (\ref{dhms}),  obtained previously
for a particular $\Phi^{\prime}$.
  
The problems discussed in this letter
can be formulated equivalently in the context of Chern-Simons
theory.  In \cite{ansar1} the complete set of LLL wavefunctions
were written down for bosons 
interacting with both a statistics transmuting
Chern-Simons field and an external magnetic field.  After a suitable
gauge transformation the wavefunctions were shown to be multivalued
and multicomponent consistent with braid group arguments.  Thus, this
is a model for anyons in an external magnetic field.  

We will follow closely Ref.\ \cite{ansar1} (We note that the notation in
Ref.\ \cite{ansar1} is different from that used here).   
Each wavefunction is specified by a
$N$ component vector of integers $\bf c$ and an integer 
$0\leq K<\gcd\left(\Phi^{\prime},N\right)$
($\gcd$ is the greatest common divisor of two numbers).
The components of $\bf c$ are ordered $0 \leq c_{1}\leq
c_{2}\ldots\leq c_{N}< s\Phi^{\prime}
\left(s\Phi^{\prime} +kN\right) \equiv\ell$  and satisfy a
constraint: 
\begin{equation}
c_{i} + \frac{k}{s\Phi^\prime}
\sum_{j=1}^{N}c_{j}=\left(s\Phi^{\prime}
+ kN\right)\left( m_{0}
+sm_{i}\right) \label{constraint}
\end{equation}
for some integers 
$m_{0}, \left\{ m_{i}\right\}$ with $0\leq m_{0}<s$.
In addition, we require that the first 
component of the vector $\bf c$, $c_{1}$,
be strictly less than $\ell /\gcd\left( N,\Phi^{\prime}\right)$.  
Given a solution $\bf c$ to the above constraint (\ref{constraint}), 
we define a new
integer $M$ as the {\em smallest}   integer such that the action
$c_{i}\rightarrow c_{i} + M\frac{\ell}
{\gcd\left(\Phi^{\prime}, N\right)} $
results in a permutation of the 
components of $\bf c$ with the components
defined $\bmod\ell$ .   
Then $K$ takes $M$ values: $n\gcd\left(\Phi^{\prime}, N\right)/M, 
n=0,\ldots ,M-1$.

To count the DHMS (denoted also as $D$)
in  the LLL  we need to count all the vectors
$\bf c$ satisfying the constraint 
(\ref{constraint}) for a given $M$, let us
call this number $d\left( M\right)$.  Then the DHMS 
 is given by 
${ D}= \sum_{M=1}^{\gcd\left(\Phi^{\prime}, N\right)}Md\left( M\right)$.
The $M$ multiplying $d\left( M\right)$ is 
due to the $M$ values taken by $K$.
We notice that $D$ can be equivalently calculated
as the {\em total} number of $\bf c$'s satisfying the constraint
(\ref{constraint}) with $0 \leq c_{1}\leq
c_{2}\ldots\leq c_{N}< s\Phi^{\prime}
\left( s\Phi^{\prime} +kN\right) \equiv\ell$ with no further
restriction on the range of $c_{1}$.  
This simplifies the problem considerably.
It immediately follows that the most general 
form for $\bf c$ is given by:
$c_{i} = s\left( c_{0} + \left( s\Phi^{\prime} 
+ kN\right)n_{i}\right) \label{sol}$, 
with $0\leq n_{1}\leq n_{2}\leq\ldots\leq n_{N}<\Phi^{\prime}\, $, 
$0\leq c_{0}< \left( s\Phi^{\prime} +kN\right)$ and 
the condition that
\begin{equation}
\frac{c_{0}+k\sum_{i=1}^{N}n_{i}}
{\Phi^\prime} \in Z \label{cond}
\end{equation}
The number of distinct vectors $\bf c$
is given by the  total number of values 
that $c_{0}$ can take times the total
number of integers $0\leq n_{1}\leq\ldots\leq n_{N}<\Phi^{\prime}$.  
The number of $c_{0}$
is simply $\left( s\Phi^{\prime} +kN\right)$, 
while the total number of $n_{i}$ is
given by the standard combinatoric result
$\left( \Phi^{\prime} +N-1\right)!/[
\left(\Phi^{\prime} -1\right)!N!]$. 
Thus the total number of vectors $\bf c$  is given by
\begin{equation}
\left( s\Phi^{\prime} +kN\right)
\frac{\left( \Phi^{\prime} +N-1\right)!}{\left(\Phi^{\prime} 
-1\right)!N!}.
\label{almost}
\end{equation}
How many of these vectors satisfy the condition (\ref{cond})?  
If $kN$ is divisible by $\Phi^{\prime}$
then for every given set of integers $\left\{ n_{i}\right\}$, 
we can find a sequence containing 
$\left(kN+s\Phi^{\prime}\right)/\Phi^{\prime}$
values of $c_{0}$ 
separated by a multiple of $\Phi^{\prime}$ 
such that condition (\ref{cond}) is
fulfilled. Therefore, $D$, the total degeneracy of LLL states, 
is given by (\ref{almost}) divided by $\Phi^{\prime}$,
which leads to our previous  
formula (\ref{dhms}) for the DHMS of LLL states.
In cases where $kN$ is not divisible by $\Phi^\prime$ we have failed
to find a rigorous proof of formula (\ref{dhms}), 
however, we have checked
it for a number of potentially anomalous cases and found that
it gives the correct degeneracy.

The formula (\ref{dhms}) for the DHMS  
is sensible for a number of cases.
Even though we have implicitly 
assumed that $\Phi^{\prime}\neq 0$ the formula
gives a sensible answer even for $\Phi^{\prime}=0$, 
namely ${ D}=k$ which
agrees with the result of \cite{ansar1}.  
For the fermionic ($k=s=1$) and 
bosonic ($k=0, s=1$) cases the standard degeneracy is reproduced. 
Following Ref.\ \cite{wu}, we  also calculated
the thermodynamics on a torus by using the formula of the DHMS 
and we found that the thermodynamics  is the same as 
that on a sphere (see also Ref.\ \cite{ouvry}).
To compare Eq.\ (\ref{dhms}) with Eq.\ (\ref{size}), 
we write $D$ as $D_1={(\Phi^{\prime}+N-1)!/[(\Phi^{\prime}-1)!N!]}$
times $D_2=(s\Phi{\prime}+kN)/\Phi^{\prime}$,
where $\Phi^{\prime}=\Phi -\alpha N$
corresponds to $d$ and $\alpha$
corresponds to $g$ in Eq.\ (\ref{core}) and Eq.\ (\ref{size}).
$D_1$ is Haldane's `choose' expression   (Eq.\ (\ref{size}))
and $D_2$ is the correction due to the center coordinate degeneracy.

We now apply the above results to the FQHE.
In the case of QH excitations of a Laughlin state
at filling $1/m$,  the effective flux $\Phi^{\prime}$ for 
the quasiparticle,
the sum of the magnetic flux and the statistical flux is equal
to the number of electrons $N_e$ in the parent state. 
The statistical parameter $\alpha$ is $1/m$. Let $N_q$ be
the number of quasiholes, we
apply formula  (\ref{dhms}) to the current case,
the DHMS of the low-energy spectrum is
${(N_e+N_q-1)!\over (N_e)!(N_q)!}(mN_e+N_q)$.

The DHMS  in the case of QE
excitations in a Laughlin state can also be worked out.
The statistical flux and the magnetic flux are now parallel,
and it is important to use hard-core boundary conditions
as in Refs.\ \cite{he,joh,sphere} to obtain the following formula. 
One part of the wavefunction
is now $\prod_{i<j} {[\theta_3(z_{i}- z_j |\tau)]}^{-1/m}
F(z_{i})_{n}$. However because of  hard-core boundary conditions,
$F(z_{i})_{n}$ should be factorized as 
$\prod_{i<j} {[\theta_3(z_{i}- z_j |\tau)]}^2 F_q$. We
note $F(z_{i})_{n}$ and $F_q(z_{i})$ have the same CCD. 
One can then derive the number of solutions
of $F_q(z_{i})$ (following the methods used previously), 
which is the DHMS  in the case of QE excitations:
${(N_e-N_q-1)!\over (N_e-2N_q)!(N_q)!}(mN_e-N_q)$.

When $N_q=0$,  the formulas of the DHMS in both cases
are equal to $m$, which is the 
degeneracy of the Laughlin state on the torus.
$m$ is the CCD of the Laughlin state, which was investigated
in detail by Haldane (the third reference in Ref.\ \cite{degen}).
 
The details of the derivations of this paper
and  the formula in the case of multispecies 
anyons with mutual statistics
on a torus will be presented  elsewhere.

DL would like to thank Prof. S. Ouvry and Prof. P. Sodano
for  stimulating discussions and encouragement,
 and  Prof. R. Iengo for discussions 
on anyon physics over the last several years.  
He also thanks CNRS for the supports
of his stay at IPN, Orsay where the work started.
AF would like to thank A. Karlhede for an enlightening discussion
on thermodynamics of anyons.

\end{document}